\documentclass[a4paper,openany, 12pt]{book}

\usepackage[text={15.5cm,23cm},centering]{geometry}

\usepackage{mathptmx}
\usepackage{graphicx}
\usepackage{chapterbib} 
\usepackage{subfigure}
\usepackage{color}
\usepackage[colorlinks=true, linkcolor=blue]{hyperref}
\usepackage[english]{babel}  

\title{The cosmic 21-cm revolution: charting the first billion years of our Universe}
\author{Andrei Mesinger}

\begin{document}

\chapter{Theoretical Framework: The Fundamentals of the 21-cm Line}

{\bf Steven R. Furlanetto, University of California, Los Angeles}

Excerpted from {\it The Cosmic 21-cm Revolution: Charting the first billion years of our Universe}, Ed. Andrei Mesinger (Bristol: IOP Publishing Ltd) AAS-IOP ebooks, \url{http://www.iopscience.org/books/aas}

\vskip 0.5in

\begin{bf}
  
We review some of the fundamental physics necessary for computing the highly-redshifted spin-flip background. We first discuss the radiative transfer of the 21-cm line and define the crucial quantities of interest. We then review the processes that set the spin temperature of the transition, with a particular focus on Wouthuysen-Field coupling, which is likely to be the most important process during and after the Cosmic Dawn. Finally, we discuss processes that heat the intergalactic medium during the Cosmic Dawn, including the scattering of Lyman-$\alpha$, cosmic microwave background, and X-ray photons.  \\
\end{bf}

Before discussing the astrophysics accessible through the 21-cm line, in this chapter we first review the basic physics required to predict the signal. 

\section{Radiative Transfer of the 21-cm Line} \label{rt-21cm}

Consider a spectral line labeled by 0 (the lower level) and 1 (the upper level). The radiative transfer equation for the specific intensity $I_\nu$ of photons at the relevant frequency is
\begin{equation}
{dI_\nu\over d\ell}={\phi(\nu) h\nu\over 4\pi}\left[n_1 A_{10} -
\left(n_0 B_{01} -n_1 B_{10}\right)I_\nu\right],
\label{rad}
\end{equation}
where $d\ell$ is a proper path length element, $\phi(\nu)$ is the line profile function, $n_i$
denotes the number density of atoms at the different levels, and $A_{ij}$ and $B_{ij}$ are the Einstein coefficients for the relevant transition (here $i$ and $j$ the initial and final states, respectively). For the 21-cm line, the line frequency is $\nu_{21} = 1420.4057$~MHz. The Einstein relations associate the radiative transition rates via $B_{10}=(g_0/g_1)B_{01}$ and $B_{10}=A_{10}(c^2/2 h\nu^3)$, where $g$ is the spin degeneracy factor of each state. For the 21-cm transition, $A_{10}=2.85\times 10^{-15} \ {\rm s^{-1}}$ and $g_1/g_0=3$.

The relative populations of hydrogen atoms in the two spin states determine the {\bf spin temperature}, $T_S$, through the relation
\begin{equation}
\left({n_1\over n_0}\right)=\left({g_1 \over g_0}\right)
\exp\left\{ {-T_*\over T_S}\right\}, 
\end{equation}
where $T_* \equiv E_{10} /k_B=68$~mK is equivalent to the transition energy $E_{10}$. In almost all physically plausible situations,  $T_\star$ is much smaller than any other temperature, including $T_S$, so all the exponentials in temperature can be Taylor expanded to leading order with high accuracy. Note, however, that $T_S$ implicitly assumes that the level populations can be described by a single temperature -- independent of each atom's velocity. In detail, velocity-dependent effects must be considered in certain circumstances \cite{hirata07}.

It is conventional to replace $I_{\nu}$ by the equivalent {\bf brightness temperature}, $T_b(\nu)$, required of a blackbody radiator (with spectrum $B_{\nu}$) such that $I_{\nu}=B_{\nu}(T_b)$. In the low frequency regime relevant to the 21 cm line, the Rayleigh-Jeans formula is an excellent approximation to the Planck curve, so $T_b(\nu)\approx I_{\nu} \, c^2/2k_B{\nu}^2$.

In this limit, the equation of radiative transfer  along a line of sight through a cloud of uniform excitation temperature $T_S$ becomes
\begin{equation}
T_b'(\nu) = T_{S}(1-e^{-\tau_{\nu}})+T_R'(\nu)e^{-\tau_{\nu}}
\label{eq:rad_trans}
\end{equation}
where $T_b'(\nu)$ is the emergent brightness measured at the cloud and at redshift $z$, the {\bf optical depth}  $\tau_\nu \equiv \int d s \, \alpha_{\nu}$ is the integral of the absorption coefficient ($\alpha_{\nu}$)  along the ray through the cloud, $T_R'$ is the brightness of the background radiation field incident on the cloud along the ray, and $s$ is the proper distance. Because of the cosmological redshift, for the 21-cm transition an observer will measure an apparent brightness at the Earth of $T_b(\nu) = T_b'(\nu_{21})/(1+z)$, where the observed frequency is $\nu=\nu_{21}/(1+z)$. Henceforth we will work in terms of these observed quantities.

The absorption coefficient is related to the Einstein coefficients via
\begin{equation}
\alpha = \phi(\nu) {h \nu \over 4 \pi} (n_0 B_{01} - n_1 B_{10}).
\end{equation}
Because all astrophysical  applications have $T_S \gg T_*$, approximately three of four atoms find themselves in the excited state ($n_0 \approx n_1/3$).  As a result, the stimulated emission correction represented by the first term is significant.  

The fundamental observable quantity is the change in brightness temperature induced by the 21-cm line by a patch of the intergalactic medium (IGM), relative to the incident radiation field. In most models that incident field is simply the cosmic microwave background, although if other sources create a low-frequency radio background at very high redshifts, or if there is a particular source behind the IGM patch along the line of sight from the observer, a larger radio background may exist.

Consider photons incident on the patch from this background. If any redshift into resonance with the 21-cm line, they can interact with the cloud -- but only for a short time, as they will redshift out of resonance as the Universe continues to expand. Thus the Hubble expansion rate sets an effective path length through the cloud, simply equal to the distance the photon travels while it remains within the line profile. The total absorption can be calculated by integrating the IGM density across this interval, in an exactly analogous procedure to the calculation of the Gunn-Peterson Lyman-$\alpha$ optical depth \cite{field59-obs, gunn65, scheuer65}. The result is
\begin{eqnarray}
\tau_{10} & = & \frac{3}{32 \pi} \, \frac{h c^3 A_{10}}{k_B T_S \nu_{10}^2} \, \frac{x_{\rm HI} n_{\rm H}}{(1+z) \, (d v_\parallel/d r_\parallel)}  \label{eq:optdepthcosmo} \\
 & \approx  & 0.0092 \, (1+\delta) \, (1+z)^{3/2}\, \frac{x_{\rm HI}}{T_S} \, \left[ \frac{H(z)/(1+z)}{d v_\parallel/d r_\parallel} \right],
\label{optdepthcosmo-approx}
\end{eqnarray}
where $n_H$ is the hydrogen number density, $x_{\rm HI}$ is the neutral fraction, $dv_\parallel/dr_\parallel$ is the velocity gradient along the line of sight (here scaled to the Hubble flow). In the second part,$T_S$ is in Kelvins, and we have scaled the density to the mean value by writing $n_H = \bar{n}_H^0 (1+z)^3 (1 + \delta)$, where $\bar{n}_H^0$ is the mean comoving density today. Note that this expression assumes a delta-function line profile, an assumption which breaks down in regimes where the peculiar velocity gradient is large. A more careful approach is required in those cases, though note that such regions are rare in most scenarios \cite{mao12}.

In most circumstances, the CMB provides the background radiation source, for which with temperature $T_\gamma(z)$. Then $T_R' = T_{\gamma}(z)$, so that  we are observing the contrast between high-redshift hydrogen clouds and the CMB.   Because the optical depth is so small, we can then expand the exponentials in equation~(\ref{eq:rad_trans}), and
\begin{eqnarray}
& T_b(\nu) & \approx \frac{T_S-T_{\gamma}(z)}{1+z}\;\tau_{\nu_0} 
\label{eq:dtbone} \\
& \approx & 9\;x_{\rm HI}(1+\delta) \, (1+z)^{1/2}\, \left[1-\frac{T_{\gamma}(z)}{T_S}\right] \, \left[ \frac{H(z)/(1+z)}{d v_\parallel/d r_\parallel} \right] \ \mbox{mK}.
\label{eq:dtb}
\end{eqnarray}
Thus $T_b < 0$ if $T_S < T_{\gamma}$, yielding an absorption signal; otherwise it appears in emission relative to the CMB. Both regimes are likely important for the high-$z$ Universe. Note that $T_b$ saturates if $T_S \gg T_{\gamma}$, but the absorption can become arbitrarily large if $T_S \ll T_{\gamma}$.  The observability of the 21 cm transition therefore hinges on the spin temperature; in the next section we will describe the mechanisms that control that factor.

Of course, the other factors -- the density, velocity, and ionization fields -- are also important to understanding the 21-cm signal. The density field evolves through cosmological structure formation, and that same evolution drives the velocity field -- both of which we will describe briefly in Chapter 3. The ionization field depends, in most scenarios, on astrophysical sources, and it will be described in detail in Chapter 2. For now, we will simply note that so long as stars drive reionization, the ``two-phase" approximation is very accurate: the mean free path of ionizing photons is so short that regions around ionizing sources are essentially fully ionized, while those outside of those H~II regions are nearly fully neutral. Thus to a good approximation, we can take $x_{\rm HI}=0$ or 1.

\section{The Spin Temperature} \label{spin-temp}

Three competing processes determine $T_S$: {\it (i)} absorption of CMB photons (as well as stimulated emission); {\it (ii)} collisions with other particles; and{\it (iii)} scattering of UV photons.  In the presence of the CMB
alone, the spin states reach thermal equilibrium ($T_S=T_{\gamma}$) on a time-scale of $\sim T_*/(T_\gamma A_{10}) = 3 \times 10^5 (1+z)^{-1}$ yr -- much shorter than the age of the Universe at all redshifts after cosmological recombination, indicating that CMB coupling establishes itself rapidly. Indeed all the relevant processes adjust on very short timescales (compared to the Hubble time) so equilibrium is an excellent approximation.

However, the other two processes break this coupling. We let $C_{10}$ and $P_{10}$ be the de-excitation rates (per atom) from collisions and UV scattering, respectively.  We also let $C_{01}$ and $P_{01}$ be the
corresponding excitation rates.  In equilibrium, the spin temperature is then
determined by
\begin{equation}
n_1 \left( C_{10} + P_{10} + A_{10} + B_{10} I_{\rm CMB} \right) = n_0 \left( C_{01} + P_{01} + B_{01} I_{\rm CMB} \right),
\label{eq:detbal}
\end{equation}
where $I_{\rm CMB}$ is the specific intensity of CMB photons at $\nu_{21}$.  With the Rayleigh-Jeans approximation, equation (\ref{eq:detbal}) can be rewritten as
\begin{equation}
T_S^{-1} = \frac{T_\gamma^{-1} + x_c T_K^{-1} + x_\alpha T_c^{-1}}{1 + x_c + x_\alpha},
\label{eq:xdefn}
\end{equation}
where $x_c$ and $x_\alpha$ are coupling coefficients for collisions and UV scattering, respectively, and $T_K$ is the gas kinetic temperature.  Here we have used the principle of detailed balance through the relation
\begin{equation}
\frac{C_{01}}{C_{10}} = \frac{g_1}{g_0} e^{-T_\star/T_K} \approx 3 \left( 1 - \frac{T_\star}{T_K} \right).
\label{eq:c01db}
\end{equation}
We have also \emph{defined} the effective color temperature of the UV radiation field $T_c$ via
\begin{equation}
\frac{P_{01}}{P_{10}} \equiv 3 \left( 1 - \frac{T_\star}{T_c} \right).
\label{eq:tcolor}
\end{equation}
In the limit in which $T_c \rightarrow T_K$ (usually a good approximation), equation~(\ref{eq:xdefn}) may be written 
\begin{equation}
1 - \frac{T_\gamma}{T_S} = \frac{x_c + x_\alpha}{1 + x_c + x_\alpha} \, \left( 1 - \frac{T_\gamma}{T_K} \right).
\label{eq:xdefn-tfac}
\end{equation}

We must now calculate $x_c$,  $x_\alpha$, and $T_c$, which we shall do in the next subsections.

\subsection{Collisional Coupling} \label{coll}

We will first consider collisional excitation and de-excitation of the hyperfine levels, which become important in dense gas.  The coupling coefficient for collisions with species $i$ is
\begin{equation}
x_c^i \equiv  \frac{C_{10}^i}{A_{10}} \, \frac{T_\star}{T_\gamma} = \frac{n_i \, \kappa_{10}^i}{A_{10}} \, \frac{T_\star}{T_\gamma},
\label{eq:xcdefn}
\end{equation}
where $\kappa_{10}^i$ is the rate coefficient for collisional spin de-excitation in collisions (with units of cm$^3$ s$^{-1}$).  The total $x_c$ is the sum over all relevant species $i$, including collisions with (1) neutral hydrogen atoms, (2) free electrons, and (3) protons.  

\begin{figure}[]
\begin{center}
\includegraphics[width=0.5\textwidth]{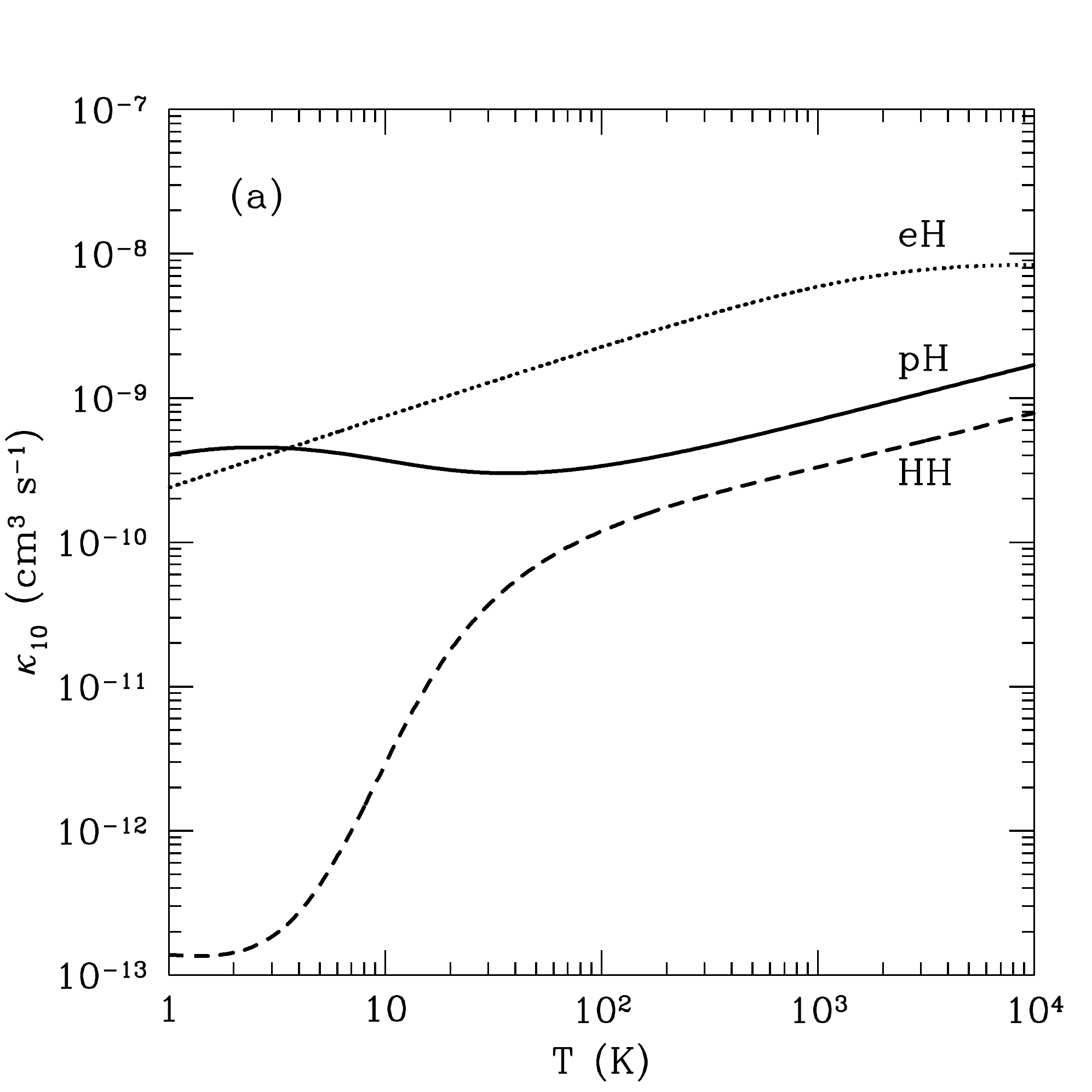}
\end{center}
\caption{De-excitation rate coefficients for H-H collisions (dashed
line), H-e$^-$ collisions (dotted line), and H-p collisions (solid
line).  Note that the net rates are also proportional to the densities
of the individual species, so H-H collisions still dominate in a
weakly-ionized medium. Reproduced from Furlanetto, S.~R. \& Furlanetto, M.~R. ``Secondary ionization and heating by fast electrons,Ó \emph{Monthly Notices of the Royal Astronomical Society}, vol. 404, pp. 1869-1878. Copyright OUP 2007.}
\label{fig:collrates}
\end{figure}

These rate coefficients can be calculated by the quantum mechanical cross sections of the relevant processes \cite{zygelman05, furl07-electron, furl07-proton}. We will not list them in detail but show the rates in Figure~\ref{fig:collrates}.  Although the atomic cross-section is small, in the unperturbed IGM collisions between neutral hydrogen atoms nearly always dominate these rates because the ionized fraction is small.  Free electrons can be important in partially ionized gas; collisions with protons are only important at the lowest temperatures.

Given the densities relevant to the IGM, collisional coupling is quite weak in a nearly neutral, cold medium.  Thus, the local density must be large in order for this process to effectively fix $T_S$. A convenient estimate of their
importance is the critical overdensity, $\delta_{\rm coll}$, at which
$x_c=1$ for H--H collisions:
\begin{equation}
1 + \delta_{\rm coll} = 0.99 \, \left[ \frac{\kappa_{10}(88 \ \mbox{K})}{\kappa_{10}(T_K)} \right] \, \left( \frac{0.023}{\Omega_b
    h^2} \right) \, \left( \frac{70}{1+z} \right)^2,
\label{eq:dcoll}
\end{equation}
where 88~K is the expected IGM temperature at $1+z=70$.\footnote{Note that this is \emph{smaller} than the CMB temperature at this time, because the IGM gas cools faster (due to adiabatic expansion) once Compton scattering becomes inefficient at $z \sim 150$.}  In the standard picture, at redshifts $z < 70$, $x_c \ll 1$ and $T_S \rightarrow T_{\gamma}$; by $z \sim 30$ the IGM essentially becomes invisible.  However, $\kappa_{10}$ is extremely sensitive to $T_K$ in this low-temperature regime.  If the Universe is somehow heated above the fiducial value, the threshold density can remain modest: $\delta_{\rm coll} \approx 1$ at $z=40$ if $T_K=300$~K.

\subsection{The Wouthuysen-Field Effect} \label{wf}

\begin{figure}[]
\begin{center}
\includegraphics[width=0.6\textwidth]{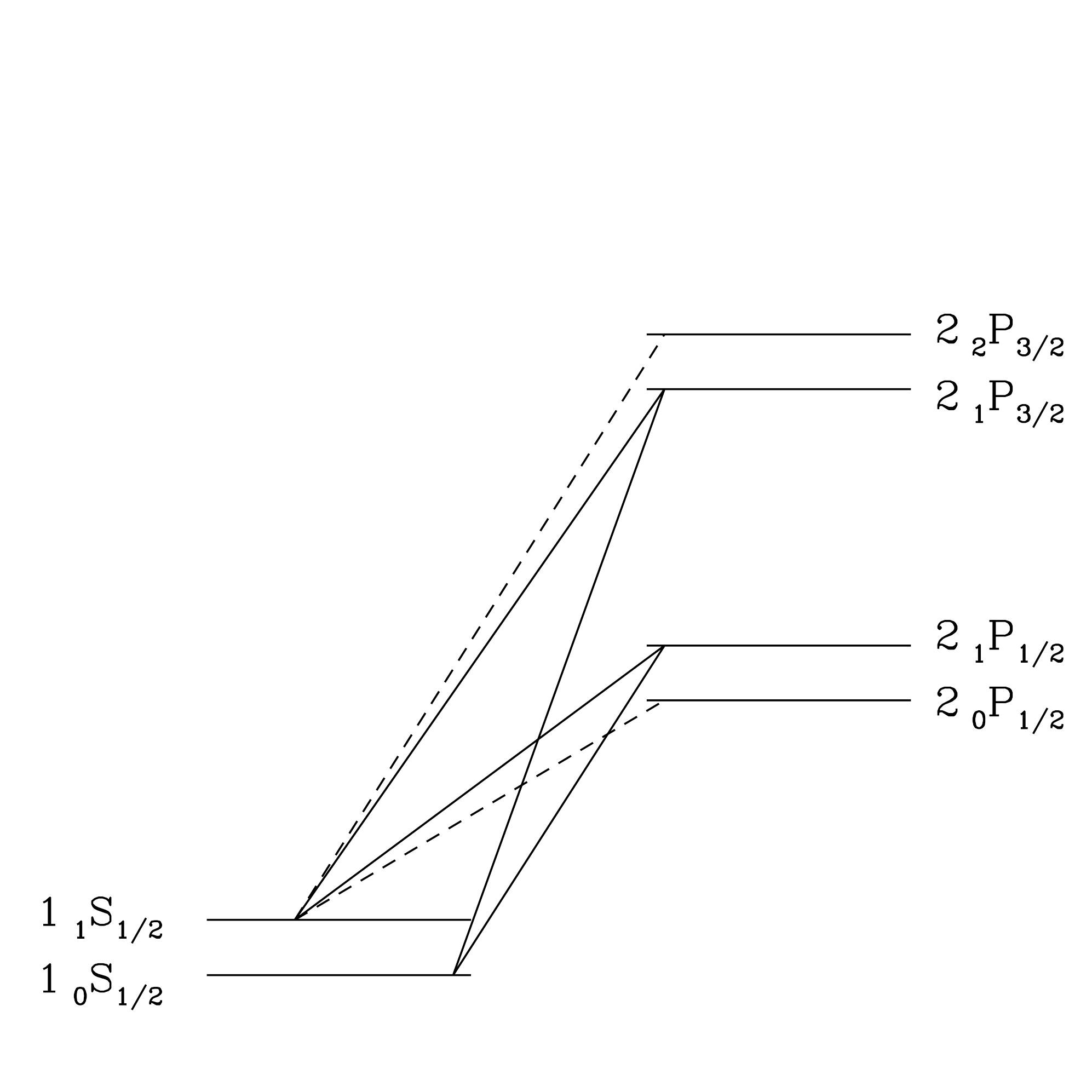}
\end{center}
\caption{Level diagram illustrating the Wouthuysen-Field effect.  We show the hyperfine splittings of the $1S$ and $2P$ levels.  The solid lines label transitions that can mix the ground state hyperfine levels, while the dashed lines label complementary allowed transitions that do not participate in mixing.  Reproduced from J.~R. Pritchard \& S.~R. Furlanetto, ``Descending from on high: Lyman-series cascades and spin-kinetic temperature coupling in the 21-cm line,Ó \emph{Monthly Notices of the Royal Astronomical Society}, vol. 367, pp. 1057-1066. Copyright OUP 2006.}
\label{fig:wf}
\end{figure}

We must therefore appeal to a different mechanism to render the 21-cm transition visible during the era of the first galaxies.  This is known as the {\bf Wouthuysen-Field mechanism} (named after the Dutch physicist Siegfried
Wouthuysen and Harvard astrophysicist George Field who first explored it \cite{wouthuysen52, field58}). Figure~\ref{fig:wf} illustrates the effect. This shows the hyperfine sub-levels of the $1S$ and $2P$ states of HI and the permitted transitions between them.  Suppose a hydrogen atom in the hyperfine singlet state absorbs a Lyman-$\alpha$ photon.  The electric dipole selection rules allow $\Delta F=0,1$ except that $F=0 \rightarrow 0$ is prohibited (here $F$ is the total angular momentum of the atom).  Thus the atom must jump to either of the central $2P$ states.  However, these same rules now allow electrons in either of these excited states to decay to the $_1S_{1/2}$ triplet level.\footnote{Here we use the notation $_F L_J$, where $L$ and $J$ are the orbital and total angular momentum of the electron.}  Thus, atoms can change hyperfine states through the absorption and spontaneous re-emission of a Lyman-$\alpha$ photon (or indeed any Lyman-series photon; see below).

The Wouthuysen-Field coupling rate depends ultimately on the total rate (per atom) at which Lyman-$\alpha$ photons scattered through the gas,
\begin{equation}
P_\alpha = 4 \pi \sigma_0 \int d \nu \, J_\nu(\nu) \phi_\alpha(\nu),
\label{eq:palpha}
\end{equation}
where $\sigma_\nu \equiv \sigma_0 \phi_\alpha(\nu)$ is the local Lyman-$\alpha$ absorption cross section, $\sigma_0 \equiv (\pi \, e^2/m_e \, c) f_{\alpha}$, $f_\alpha=0.4162$ is the oscillator strength of the Lyman-$\alpha$ transition, $\phi_\alpha(\nu)$ is the Lyman-$\alpha$ absorption profile, and $J_\nu$ is the angle-averaged specific intensity of the background radiation field.\footnote{By convention, we use the specific intensity in units of photons cm$^{-2}$ Hz$^{-1}$ s$^{-1}$ sr$^{-1}$ here, which is conserved during the expansion of the Universe (whereas a definition in terms of energy instead of photon number is subject to redshifting).}   

Transitions to higher Lyman-$n$ levels have similar effects \cite{hirata06, pritchard06}. Suppose that a UV photon redshifts into the Lyman-$n$ resonance as it travels through the IGM.  After absorption, it can either scatter (by the electron decaying directly to the ground state) or cascade through a series of intermediate levels and produce a sequence of photons.  The direct decay probability for any level is $\sim 0.8$, so a Lyman-$n$ photon will typically scatter $N_{\rm scatt} \approx (1-P_{nP\rightarrow1S})^{-1} \sim 5$ times before instead initiating a decay cascade.  In contrast, Lyman-$\alpha$ photons scatter hundreds of thousands of times before being destroyed, usually be redshifting all the way across the (very wide) Lyman-$\alpha$ profile.  As a result, coupling from the direct scattering of Lyman-$n$ photons is suppressed compared to Lyman-$\alpha$ by a large factor.

\begin{figure}[]
\begin{center}
\includegraphics[width=0.6\textwidth]{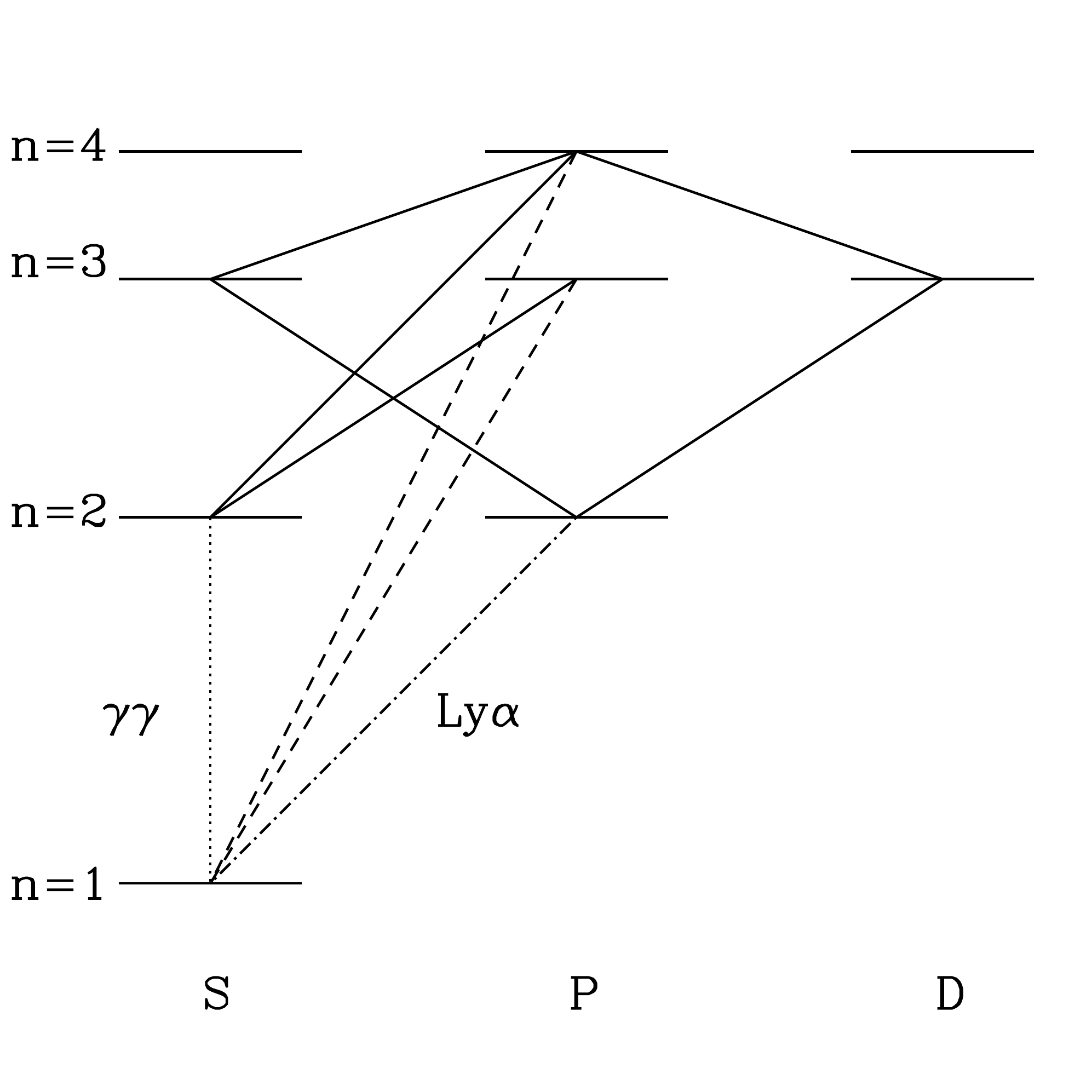}
\end{center}
\caption{Decay chains for Lyman-$\beta$ and Lyman-$\gamma$ excitations.  We show
  Lyman-$n$ transitions by dashed curves, Lyman-$\alpha$ by the dot-dashed
  curve, cascades by solid curves, and the forbidden $2S \rightarrow
  1S$ transition by the dotted curve. Reproduced from J.~R. Pritchard \& S.~R. Furlanetto, ``Descending from on high: Lyman-series cascades and spin-kinetic temperature coupling in the 21-cm line,Ó \emph{Monthly Notices of the Royal Astronomical Society}, vol. 367, pp. 1057-1066. Copyright OUP 2006.}
\label{fig:lygamma}
\end{figure}

However, Lyman-$n$ photons can still be important because of their cascade products, as shown in Figure~\ref{fig:lygamma}.  Following Lyman-$\beta$ absorption, the only permitted decays are to the ground state (regenerating a Lyman-$\beta$ photon and starting the process again) or to the $2S$ level.  The H$\alpha$ photon produced in the $3P \rightarrow 2S$ transition (and indeed any photon produced in a decay to an excited state) escapes to infinity. Thus the atom will eventually find itself in the $2S$ state, which decays to the ground state via a forbidden two photon process with $A_{2S\rightarrow1S}=8.2$~s$^{-1}$.  These photons will also escape to infinity, so coupling from Lyman-$\beta$ photons can be completely neglected.\footnote{In a medium with very high number density, atomic collisions can mix the two angular momentum states, but that process is unimportant in the IGM.}

But now consider excitation by Lyman-$\gamma$, also shown in Figure~\ref{fig:lygamma}.  This can cascade (through $3S$ or $3D$) to the $2P$ level, in which case the original Lyman-$n$ photon is ``recycled'' into a Lyman-$\alpha$ photon, which then scatters many times through the IGM.  Thus, the key quantity for determining the coupling induced by Lyman-$n$ photons is the fraction $f_{\rm rec}(n)$ of cascades that terminate in Lyman-$\alpha$ photons.  Our discussion in the previous paragraph shows that $f_{\rm rec}(n=3)$ vanishes, but detailed quantum mechanical calculations show that the higher states all have $f_{\rm rec} \sim 1/3$ \cite{hirata06, pritchard06}. 

Focusing again on the Lyman-$\alpha$ photons themselves, we must relate the total scattering rate $P_\alpha$ to the indirect de-excitation rate $P_{10}$ \cite{field58, meiksin00}. Let us first label the $1S$ and $2P$ hyperfine levels a--f, in order of increasing energy, and let $A_{ij}$ and $B_{ij}$ be the spontaneous emission and absorption coefficients for transitions between these levels.  We write the background intensity at the frequency corresponding to the $i \rightarrow j$ transition as $J_{ij}$.  Then
\begin{equation}
P_{01} \propto B_{\rm ad} J_{\rm ad} \frac{A_{\rm db}}{A_{\rm da} + A_{\rm db}} + B_{\rm ae} J_{\rm ae} \frac{A_{\rm eb}}{A_{\rm ea} + A_{\rm eb}}.\label{eq:psum}
\end{equation}
The first term contains the probability for an a$\rightarrow$d transition ($B_{\rm ad} J_{\rm ad}$), together with the probability for the subsequent decay to terminate in state b; the second term is the same for transitions to and from state e (see Figure~\ref{fig:wf}).  Next we need to relate each$A_{ij}$ to the total spontaneous decay rate from the $2P$ level, $A_\alpha = 6.25 \times 10^8$~Hz, the total Lyman-$\alpha$ spontaneous emission rate.  This can be accomplished using a sum rule stating that the sum of decay intensities ($g_i A_{ij}$) for transitions from a
given $nFJ$ to all the $n' J'$ levels (summed over $F'$) is proportional to $2F+1$, which implies that the relative strengths of the permitted transitions are then $(1,\,1,\,2,\,2,\,1,\,5)$, where we have ordered the lines by (initial, final) states (bc, ad, bd, ae, be, bf).  With our assumption that the background radiation field is constant across the individual hyperfine lines, we find $P_{10} = (4/27) P_\alpha$ \cite{meiksin00}.

The coupling coefficient $x_\alpha$ is then
\begin{equation}
x_\alpha = \frac{4 P_\alpha}{27 A_{10}} \, \frac{T_\star}{T_{\gamma}} \equiv S_\alpha \frac{J_\alpha}{J_\nu^c}.
\label{eq:xalpha}
\end{equation}
The second part evaluates $J_\nu$ ``near" line center and sets $J_\nu^c \equiv 1.165 \times 10^{-10} [(1+z)/20]$~photons cm$^{-2}$ sr$^{-1}$ Hz$^{-1}$ s$^{-1}$.   $S_\alpha$ s a correction factor that accounts for (complicated) radiative transfer effects in the intensity near the line center (see below).  The coupling threshold $J_\nu^c$ for $x_\alpha = S_\alpha$ can also be written in terms of the number of Lyman-$\alpha$ photons per hydrogen atom in the Universe, which we denote $\tilde{J}_\nu^c = 0.0767 \, [(1+z)/20]^{-2}$.  This threshold is relatively easy to achieve in practice.

To complete the coupling calculation, we must determine $T_c$ and the correction factor $S_\alpha$.  The former is the effective temperature of the UV radiation field, defined in equation~(\ref{eq:tcolor}), and is determined by the shape of the photon spectrum at the Lyman-$\alpha$ resonance. The effective temperature of the radiation field \emph{must} matter, because the energy deficit between the different hyperfine splittings of the Lyman-$\alpha$ transition (labeled bc, ad, etc. above) implies that the mixing process is sensitive to the gradient of the radiation spectrum near the Lyman-$\alpha$ resonance.  More precisely, the procedure described after equation~(\ref{eq:psum}) yields
\begin{equation}
\frac{P_{01}}{P_{10}} = \frac{g_1}{g_0} \, \frac{n_{\rm ad} + n_{\rm ae}}{n_{\rm bd} + n_{\rm be}} \approx 3 \left( 1 + \nu_0 \frac{d \ln n_\nu}{d \nu} \right),
\label{eq:tcrad1}
\end{equation}
where $n_\nu = c^2 \, J_\nu/2 \nu^2$ is the photon occupation number.
Thus, by comparison to equation~(\ref{eq:tcolor}) we find
\begin{equation}
\frac{h}{k_B T_c} = - \frac{d \ln n_\nu}{d \nu}.
\label{eq:tcrad}
\end{equation}

A simple argument shows that $T_c \approx T_K$ \cite{field59-ts}: so long as the medium is extremely optically thick, the enormous number of Lyman-$\alpha$ scatterings forces the Lyman-$\alpha$ profile to be a blackbody of temperature $T_K$ near the line center.  This condition is easily fulfilled in the high-redshift IGM, where $\tau_\alpha \gg 1$.  In detail, atomic recoils during scattering tilt the spectrum to the red and are primarily responsible for establishing this equilibrium \cite{field59-res}.  

The physics of the Wouthuysen-Field effect are actually much more complicated than naively expected because scattering itself modifies the shape of $J_\nu$ near the Lyman-$\alpha$ resonance \cite{chen04}. In essence, the spectrum must develop an absorption feature because of the increased scattering rate near the Lyman-$\alpha$ resonance. Photons lose energy at a fixed rate by redshifting, but each time they scatter they also lose a small amount of energy through recoil.  Momentum conservation during each scattering slightly decreases the frequency of the photon.  The strongly enhanced scattering rate near line center means that photons ``flow" through
that region more rapidly than elsewhere (where only the cosmological redshift applies), so the amplitude of the spectrum must be smaller.  Meanwhile, the scattering in such an optically thick medium also causes photons to diffuse away from line center, broadening the feature well beyond the nominal line width.

If the fractional frequency drift rate is denoted by ${\mathcal A}$, continuity requires $n_\nu {\mathcal A}=$~constant. Because ${\mathcal A}$ increases near resonance, the number density must fall.  On average, the energy loss (or gain) per scattering is \cite{chen04}
\begin{equation}
\frac{\Delta E_{\rm recoil}}{E} = \frac{h \nu}{m_p c^2} \, \left(1 - \frac{T_K}{T_c} \right),
\label{eq:recoil-loss}
\end{equation}
where the first factor comes from recoil off an isolated atom and the second factor corrects for the distribution of initial photon energies; the energy loss vanishes when $T_c = T_K$, and when $T_c < T_K$, the gas is heated by the scattering process.

To compute $S_\alpha$, we must calculate the photon spectrum near Lyman-$\alpha$.  We begin with the radiative
transfer equation in an expanding universe (written in comoving coordinates, and again using units of ~photons cm$^{-2}$ sr$^{-1}$ Hz$^{-1}$ s$^{-1}$ for $J_\nu$:
\begin{equation}
\frac{1}{c n_H \sigma_0} \, \frac{\partial J_\nu}{\partial t} = -\phi_\alpha(\nu) \, J_\nu + H \nu_\alpha \, \frac{\partial J_\nu}{\partial \nu} + \int d \nu' \, R(\nu,\nu') \, J_{\nu'} + C(t) \psi(\nu).
\label{eq:rt21}
\end{equation}
The first term on the right-hand side describes absorption, the second describes redshifting due to the Hubble flow, and the third accounts for re-emission following absorption.  $R(\nu,\nu')$ is the ``redistribution function" that specifies the frequency of an emitted photon, which depends on the relative momenta of the absorbed and
emitted photons as well as the absorbing atom. The last term accounts for the injection of new photons (via, e.g., radiative cascades that result in Lyman-$\alpha$ photons): $C$ is the rate at which they are produced and $\psi(\nu)$ is their frequency distribution.

The redistribution function $R$ is the difficult aspect of the problem, but it can be simplified if the frequency change per scattering (typically of order the absorption line width) is ``small."  In that case, we can expand $J_{\nu'}$ to second order in $(\nu-\nu')$ and rewrite equation~(\ref{eq:rt21}) as a diffusion problem in frequency.
The steady-state version of equation (\ref{eq:rt21}) becomes, in this so-called {\it Fokker-Planck} approximation, \cite{chen04}
\begin{equation}
\frac{d}{d x} \left( - {\mathcal A} \, J + {\mathcal D} \, \frac{d J}{d x} \right) + C \psi(x) = 0,
\label{eq:fokker}
\end{equation}
where $x \equiv (\nu-\nu_\alpha)/\Delta \nu_D$, $\Delta \nu_D$ is the Doppler width of the absorption profile, ${\mathcal A}$ is the frequency drift rate, and ${\mathcal D}$ is the diffusivity.  The Fokker-Planck approximation is valid so long as (i) the frequency change per scattering ($\sim \Delta \nu_D$) is smaller than the width of any spectral features, and either (iia) the photons are outside the line core where the Lyman-$\alpha$ line profile is slowly changing, or (iib) the atoms are in equilibrium with $T_c \approx T_K$.

Solving for the spectrum including scattering thus reduces to specifying ${\mathcal A}$ and ${\mathcal D}$.  The drift involves the Hubble flow, which sets ${\mathcal A}_H= - \tau_\alpha^{-1}$, where $\tau_\alpha$ is the Gunn-Peterson optical depth for the Lyman-$\alpha$ line \cite{gunn65, scheuer65}:
\begin{equation}
\tau_{\alpha} = \frac{\chi_\alpha \, n_{\rm HI}(z) \, c}{H(z) \nu_\alpha} \approx 3 \times 10^5 \, x_{\rm HI} \, \left( \frac{1+z}{7} \right)^{3/2}.
\label{eq:taugp}
\end{equation}
Because it is uniform, the Hubble flow does not introduce any diffusion. The remaining terms come from $R$ and
incorporate all the physical processes relevant to energy exchange in scattering.  The drift from recoil causes \cite{hirata06}
\begin{eqnarray}
{\mathcal D}_{\rm scatt} & = & \phi_\alpha(x)/2,
\label{eq:Dkin} \\
{\mathcal A}_{\rm scatt} & = &  -(\eta - x_0^{-1} ) \phi_\alpha(x),
\label{eq:Akin}
\end{eqnarray}
where $x_0 \equiv \nu_\alpha/\Delta \nu_D$ and $\eta \equiv (h \nu_\alpha^2)/(m_p c^2 \Delta \nu_D)$.  The latter is the recoil parameter measuring the average loss per scattering in units of the Doppler width.  The small energy defect between the hyperfine levels provides another source of slow energy exchange \cite{hirata06} and can be incorporated into the scattering in nearly the same way as recoil.

We can now solve equation~(\ref{eq:fokker}) once we choose the boundary conditions, which essentially correspond to the input photon spectrum (ignoring scattering) and the source function.  Because the
frequency range of interest is so narrow, two cases suffice: a flat input spectrum (which approximately describes photons that redshift through the Lyman-$\alpha$ resonance, regardless of the initial source spectrum)
and a step function, where photons are ``injected" at line center (through cascades or recombinations) and redshift away.  In either case, the first integral over $x$ in equation~(\ref{eq:fokker}) is trivial. At high temperatures where spin flips are unimportant to the overall energy exchange, we can write
\begin{equation}
\phi \frac{d J}{d x} + 2 \{ [\eta - (x + x_0)^{-1}] \phi + \tau_\alpha^{-1} \} J = 2 K / \tau_\alpha.
\label{eq:fokk-simp}
\end{equation}
The integration constant $K$ equals $J_\infty$, the flux far from resonance, both for photons that redshift into the line and for injected photons at $x<0$ (i.e., redward of line center); it is zero for injected photons at $x>0$.  

The formal analytic solution, when $K \neq 0$, is most compactly written in terms of $\delta_J \equiv (J_\infty -
J)/J_\infty$ \cite{chen04}:\footnote{Here we assume the gas has a sufficiently high temperature that the different hyperfine sub-transitions can be treated as one \cite{hirata06}.}
\begin{equation}
\delta_J(x) = 2 \eta \int_0^\infty d y \exp \left[ - 2 \{ \eta - (x+x_0)^{-1}\} y - {2 \over \tau_\alpha} \int_{x-y}^{x} \frac{d x'}{\phi_\alpha(x')} \right].
\label{eq:dj-soln}
\end{equation}
(An analogous form also exists for photons injected at line center.) The full problem, including the intrinsic Voigt profile of the Lyman-$\alpha$ line, must be solved numerically, but including only the Lorentzian wings from natural broadening allows a simpler solution \cite{furl06-lyheat}.  Fortunately, this assumption is quite accurate in the most interesting regime of  $T_K <  1000$~K.

\begin{figure}[]
\begin{center}
\includegraphics[width=0.6\textwidth]{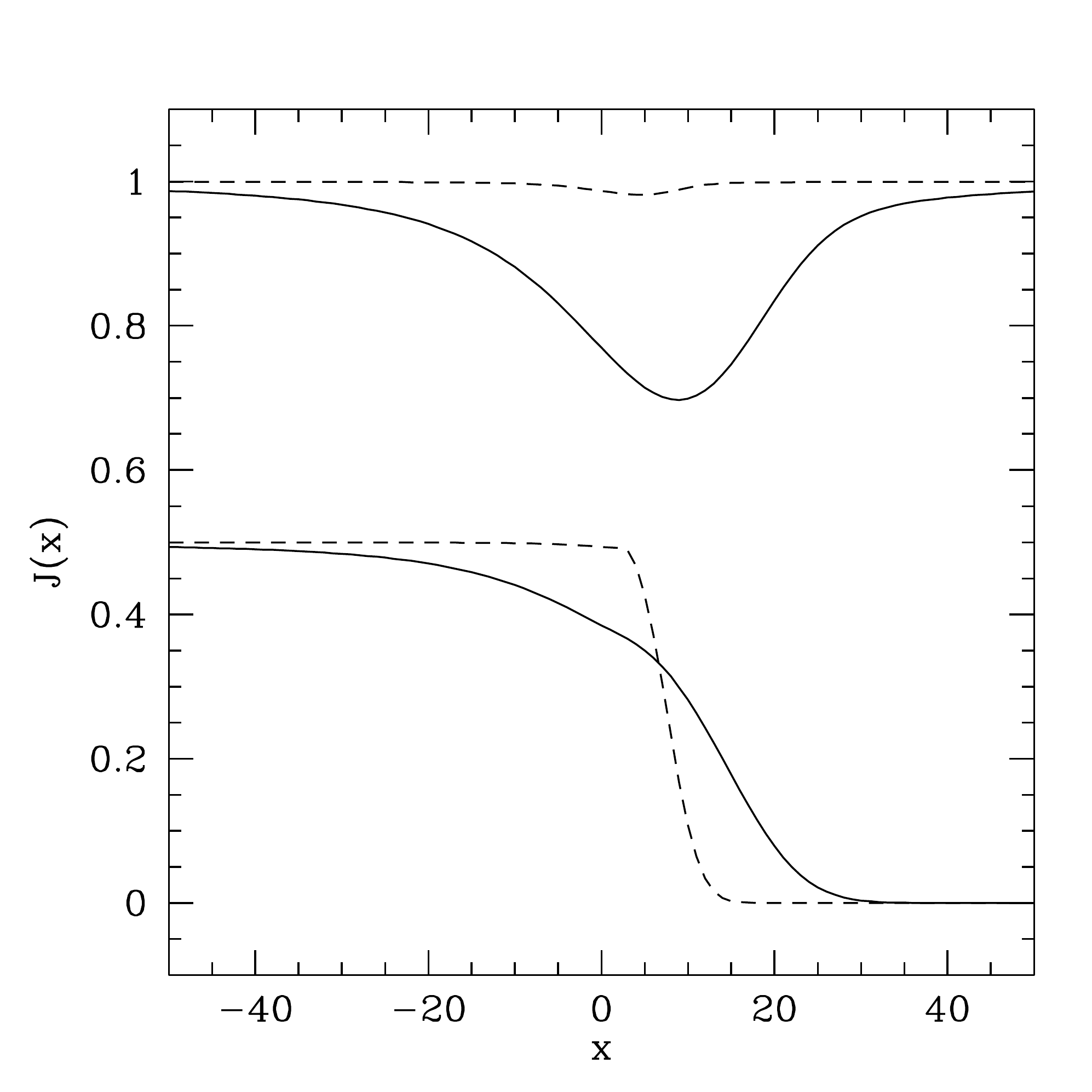}
\end{center}
\caption{Background radiation field near the Lyman-$\alpha$ resonance at $z=10$;
$x \equiv (\nu-\nu_\alpha)/\Delta \nu_D$ is the normalized deviation
from line center, in units of the Doppler width.  The upper and lower sets are for continuous photons
and photons injected at line center, respectively.  (The former are
normalized to $J_\infty$; the latter have arbitrary normalization.)
The solid and dashed curves take $T_K=10$ and $1000$~K,
respectively.  Reproduced from S.~R. Furlanetto \& J.~R., Pritchard, ``The scattering of Lyman-series photons in the intergalactic medium," \emph{Monthly Notices of the Royal Astronomical Society}, vol. 372, pp. 1093-1103. Copyright OUP 2006. }
\label{fig:lyshape}
\end{figure}

The crucial aspect of equation~(\ref{eq:dj-soln}) is that (as expected from the qualitative argument) an absorption feature appears near the line center, with its depth roughly proportional to $\eta$, our recoil parameter.  The feature is more significant when $T_K$ is small (because in that case the average effect of recoil is large). Figure~\ref{fig:lyshape} shows some example spectra (both for a continuous background and for photons injected at line center).

Usually, the most important consequence is the suppression of the radiation spectrum at line center compared to the assumed initial condition.  This decreases the total scattering rate of Lyman-$\alpha$ photons (and hence the Wouthuysen-Field coupling), with the suppression factor (defined in equation~\ref{eq:xalpha}) as \cite{chen04} 
\begin{equation}
S_\alpha = \int_{-\infty}^\infty d x \, \phi_\alpha(x)\, J(x) \approx
[1 - \delta_J(0)] \le 1,
\label{eq:salpha-defn}
\end{equation}
where the second equality follows from the narrowness of the line profile.  Again, the Lorentzian wing approximation turns out to be an excellent one; when $T_K \gg T_\star$, the suppression is \cite{furl06-lyheat}
\begin{equation} S_\alpha \sim \exp \left[ -0.803 \left({T_K\over
1~{\rm K}}\right)^{-2/3} \left( \frac{\tau_\alpha}{10^{6}}
\right)^{1/3} \right].
\label{eq:salpha-approx}
\end{equation}
Note that this form applies to both photons injected at line center as well as those that redshift in from infinity.  As we can see in Figure~\ref{fig:lyshape}, the suppression is most significant in cool gas.

\section{Heating of the Intergalactic Medium} 

We have seen that both collisions and the Wouthuysen-Field effect couple the spin temperature to the kinetic temperature of the gas. The 21-cm brightness temperature therefore depends on processes that heat the \emph{neutral} IGM. (Note that photoionization heating is likely the most important mechanism in setting the IGM temperature, because that process typically heats the gas to $T \sim 10^4$~K. However, by definition that process only occurs when ionization is significant -- and, in standard reionization scenarios, where $x_{\rm HI} \approx 0$ so that the 21-cm signal vanishes.) We will review several such mechanisms in this section.

\subsection{The Lyman-$\alpha$ Background}

The photons that trigger Lyman-$\alpha$ coupling exchange energy with the IGM, through the recoil in each scattering event. The typical energy exchange per scattering is small (see eq. \ref{eq:recoil-loss}), but the scattering rate is extremely large.  If the net heating rate per atom followed the naive expectation, $\sim
P_\alpha \times (h \nu_\alpha)^2/m_p c^2$, the kinetic temperature would surpass $T_\gamma$ soon after Wouthuysen-Field coupling becomes efficient.

However, the details of radiative transfer radically change these expectations \cite{chen04}.  In a static medium, the energy exchange \emph{must} vanish in equilibrium even though scattering continues at nearly the same rate.
Scattering induces an asymmetric absorption feature near $\nu_\alpha$ (Figure~\ref{fig:lyshape}) whose shape depends on the combined effects of atomic recoils and the scattering diffusivity.  In equilibrium, the latter exactly counterbalances the former.  

If we removed scattering, the absorption feature would redshift away as the Universe expands. Thus, the energy exchange rate from scattering must simply be that required to maintain the feature in place.  For photons redshifting into resonance, the absorption trough has total energy 
\begin{equation}
\Delta u_\alpha = (4\pi/c) \int (J_\infty - J_\nu) h \nu d \nu,
\end{equation}
where $J_\infty$ is the input spectrum, and we note that the $h \nu$ factor converts from our definition of specific intensity (which counts photons) to energy.  The radiation background loses $\epsilon_\alpha = H \Delta u_\alpha$ per
unit time through redshifting; this energy goes into heating the gas.
Relative to adiabatic cooling by the Hubble expansion, the fractional
heating amplitude is
\begin{eqnarray}
\frac{2}{3} \, \frac{\epsilon_\alpha}{k_B T_K n_H H(z) } & = & \frac{8 \pi}{3} \, \frac{h \nu_\alpha}{k_B T_K} \, \frac{J_\infty \, \Delta \nu_D}{c n_H} \, \int_{-\infty}^{\infty} d x \delta_J(x) \label{eq:epsalpha}
\\
& \approx & \frac{0.80}{T_K^{4/3}} \, \frac{x_\alpha}{S_\alpha} \left( \frac{10}{1+z} \right),
\label{eq:lyaheat}
\end{eqnarray}
Here we have evaluated the integral for the continuum photons that
redshift into the Lyman-$\alpha$ resonance; the ``injected" photons
actually cool the gas slightly.  The net energy exchange when
Wouthuysen-Field coupling becomes important (at $x_\alpha \sim
S_\alpha$) is therefore just a fraction of a degree, and in practice gas heating through Lyman-$\alpha$ scattering
is generally unimportant \cite{chen04,furl06-lyheat}.

Fundamentally, Lyman-$\alpha$ heating is inefficient because scattering diffusivity cancels the effects of recoil.  From Figure~\ref{fig:lyshape}, we see that the background spectrum is weaker on the blue side of the line than on the red.  The scattering process tends to move the photon toward line center, with the extra energy deposited in or extracted from the gas.  Because more scattering occurs on the red side, this tends to transfer energy from the gas back to the photons, mostly canceling the energy obtained through recoil.

\subsection{The Cosmic Microwave Background} 

The previous section shows that, when considered as a two-level process that acts in isolation, Lyman-$\alpha$ scattering has only a slight effect on the gas temperature. However, in reality this Lyman-$\alpha$ scattering always occurs in conjunction with scattering of CMB photons within the 21-cm transition. The combination leads to an enhanced heating rate \cite{venumadhav18}.

In essence, the process works as follows. The CMB photons scatter through the hyperfine levels of HI to heat those atoms above their expected temperature (determined in this simple case by adiabatic cooling). Meanwhile, Lyman-$\alpha$ photons scatter through the gas as well. As they do so, they mix the hyperfine levels of the HI ground state, as depicted in Figure~\ref{fig:wf} -- this is the Wouthuysen-Field effect. CMB scattering continues to heat the hyperfine level populations during the Lyman-$\alpha$ scattering, which then sweeps up this extra energy and ultimately deposits it as thermal energy through the net recoil effect.

We can estimate the energy available to this heating mechanism by considering the CMB energy reservoir \cite{venumadhav18}. The CMB energy density at the 21-cm transition is $u_\nu = (4 \pi/c) B_\nu \approx 8 \pi (\nu_{21}^2/c^3) k_B T_\gamma$. Over a redshift interval $\Delta z=1$, the total energy that redshifts through the line is $u_\nu \Delta \nu \approx 8 \pi (\nu_{21}/c)^3 k_B T_\gamma / (1+z)^2$. However, only a fraction $\tau_{10}$ actually interacts with the line. If all of this energy is used for heating, the temperature change per H atom would be
\begin{equation}
\Delta T_{\rm CMB-Ly\alpha} \approx \tau_{10} {u_\nu \Delta_\nu \over (3/2) n_H} \approx 5 x_{\rm HI} \  \left( {1 + z \over 20} \right)^{-1/2} \left( {10 \ {\rm K} \over T_S} \right) \ {\rm K}.
\label{eq:cmb-lya-heat}
\end{equation}

A more detailed calculation of the heating rate shows that it is somewhat slower, but it does amplify the effect of the Lyman-$\alpha$ heating alone by a factor of several \cite{venumadhav18}. In standard models of the early radiation backgrounds, the correction is still relatively modest, but it is not negligible. For example, in the fiducial model considered by \cite{venumadhav18}, the Lyman-$\alpha$ heating on its own modifies $T_K$ by $\sim 1$--5\%, but with the CMB scattering included the effect is $\sim 9$--15\%. Additionally, the CMB scattering can be enhanced in some exotic physics models that decrease the spin temperature substantially.

\subsection{The X-ray Background} \label{xrayheat}

Because they have relatively long mean free paths, X-rays from galaxies and quasars are likely to be the most important heating agent for the low-density IGM  \cite{madau97}.  In particular, photons with $E>1.5 x_{\rm HI}^{1/3} [(1+z)/10]^{1/2}$~keV have mean free paths exceeding the Hubble length \cite{oh01}.  Lower-energy X-rays will be absorbed in the IGM, depositing much of their energy as heat, as will a fraction of higher-energy X-rays.

X-rays heat the IGM gas by first photoionizing a hydrogen or helium atom.  The resulting ``primary'' electron retains most of the photon energy (aside from that required to ionize it) as kinetic energy, which it must then distribute to the general IGM through three main channels: (1) collisional ionizations, which produce more secondary electrons that themselves scatter through the IGM, (2) collisional excitations of HeI (which produce photons capable of ionizing HI) and HI (which produces a Lyman-$\alpha$ background), and (3) Coulomb collisions with free electrons (which distributes the kinetic energy .  The relative cross-sections of these processes determines what fraction of the X-ray energy goes to heating ($f_{\rm heat}$), ionization ($f_{\rm ion}$), and excitation ($f_{\rm excite}$); clearly it depends on both the ionized fraction $x_i$ and the input photon energy.  Through these scatterings, the primary photoelectrons, with $T \sim 10^6$~K, rapidly cool to energies just below the Lyman-$\alpha$ threshold, $<10$~eV, and thus equilibrate with the other IGM electrons.  After that, the electrons and neutrals equilibrate through elastic scattering on a timescale $t_{\rm eq} \sim 5 [10/(1+z)]^3$~Myr.  Because $t_{\rm eq} \ll H(z)^{-1}$, the assumption of a single temperature fluid is an excellent one.

\begin{figure}[]
\begin{center}
\includegraphics[width=0.4\textwidth]{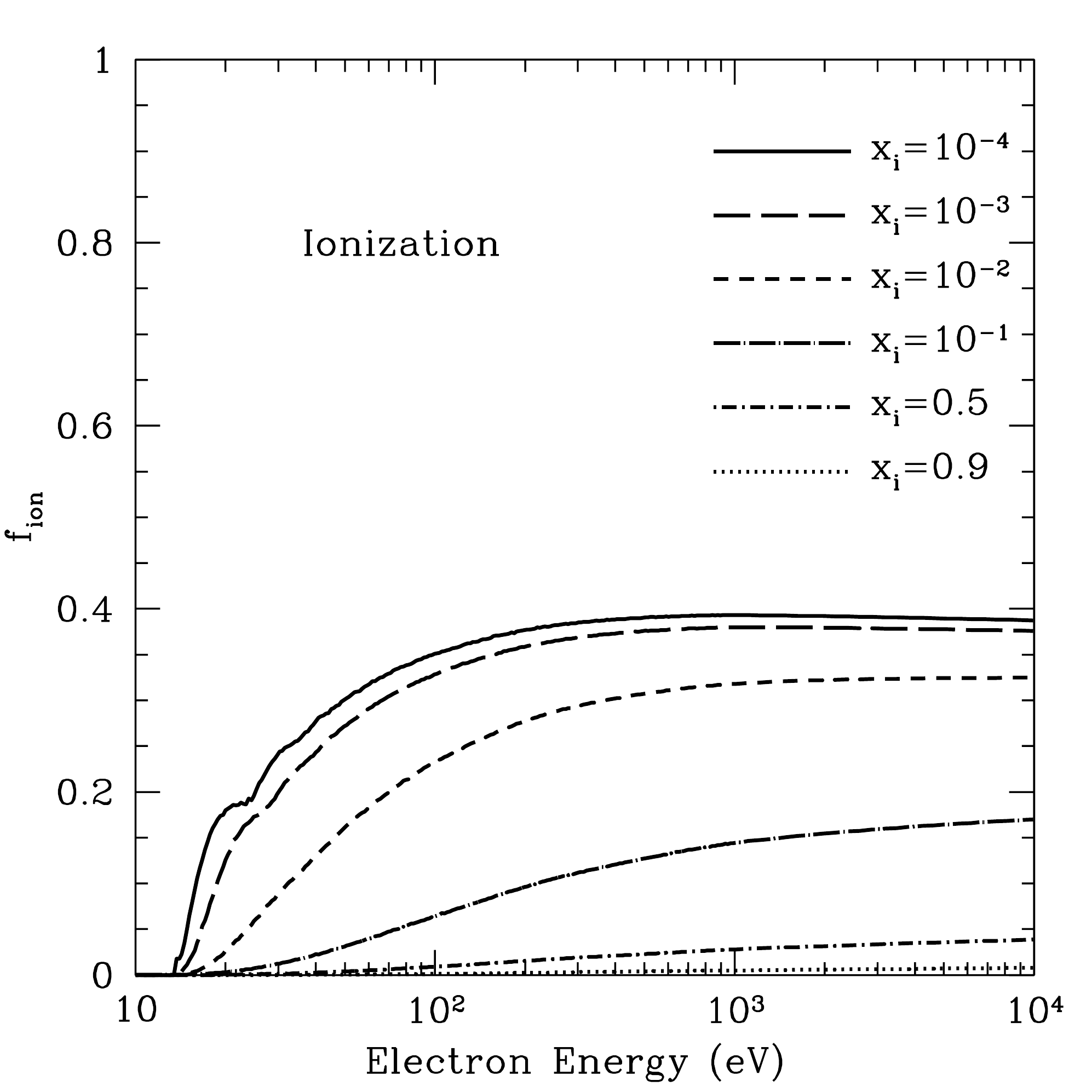} \includegraphics[width=0.4\textwidth]{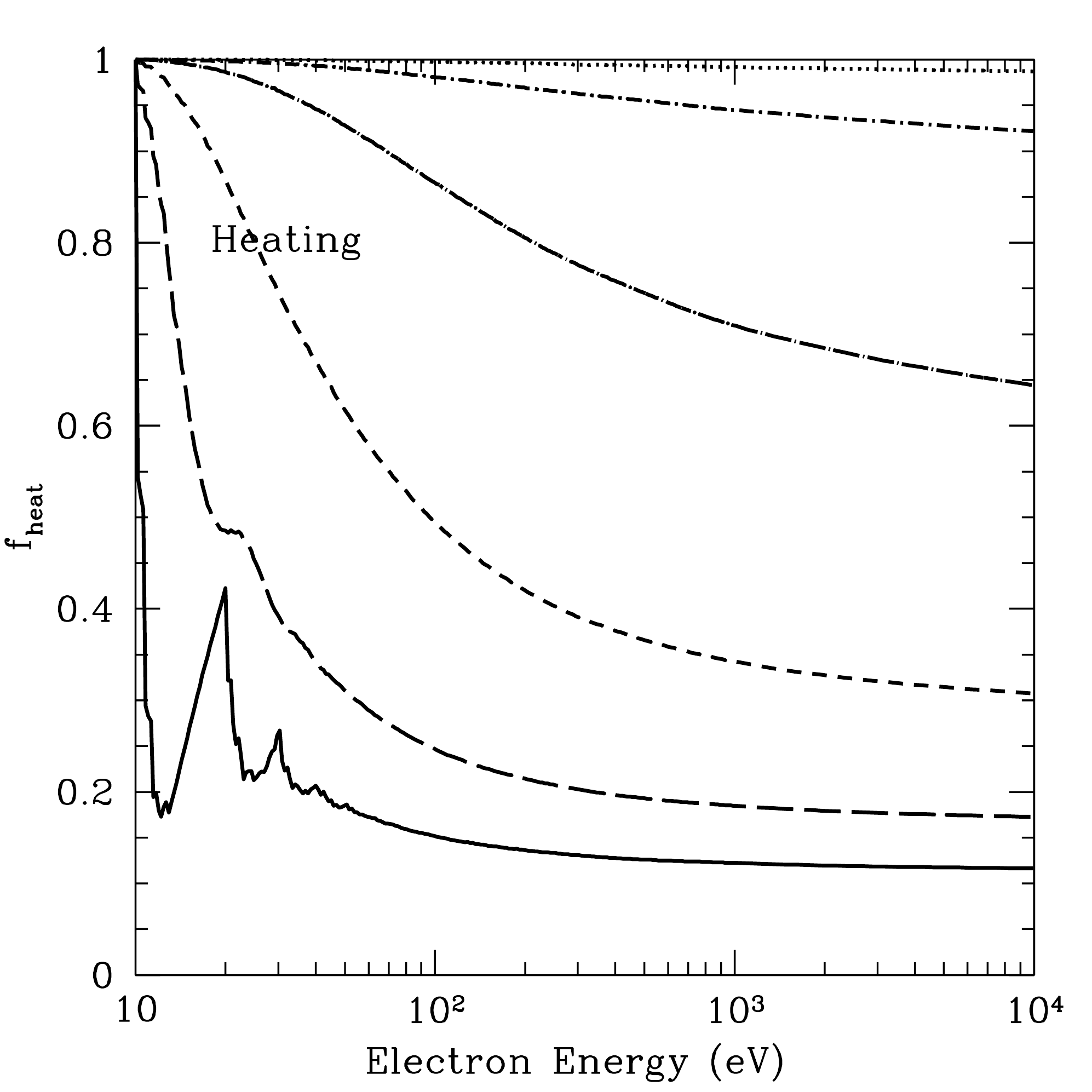} \\
\includegraphics[width=0.4\textwidth]{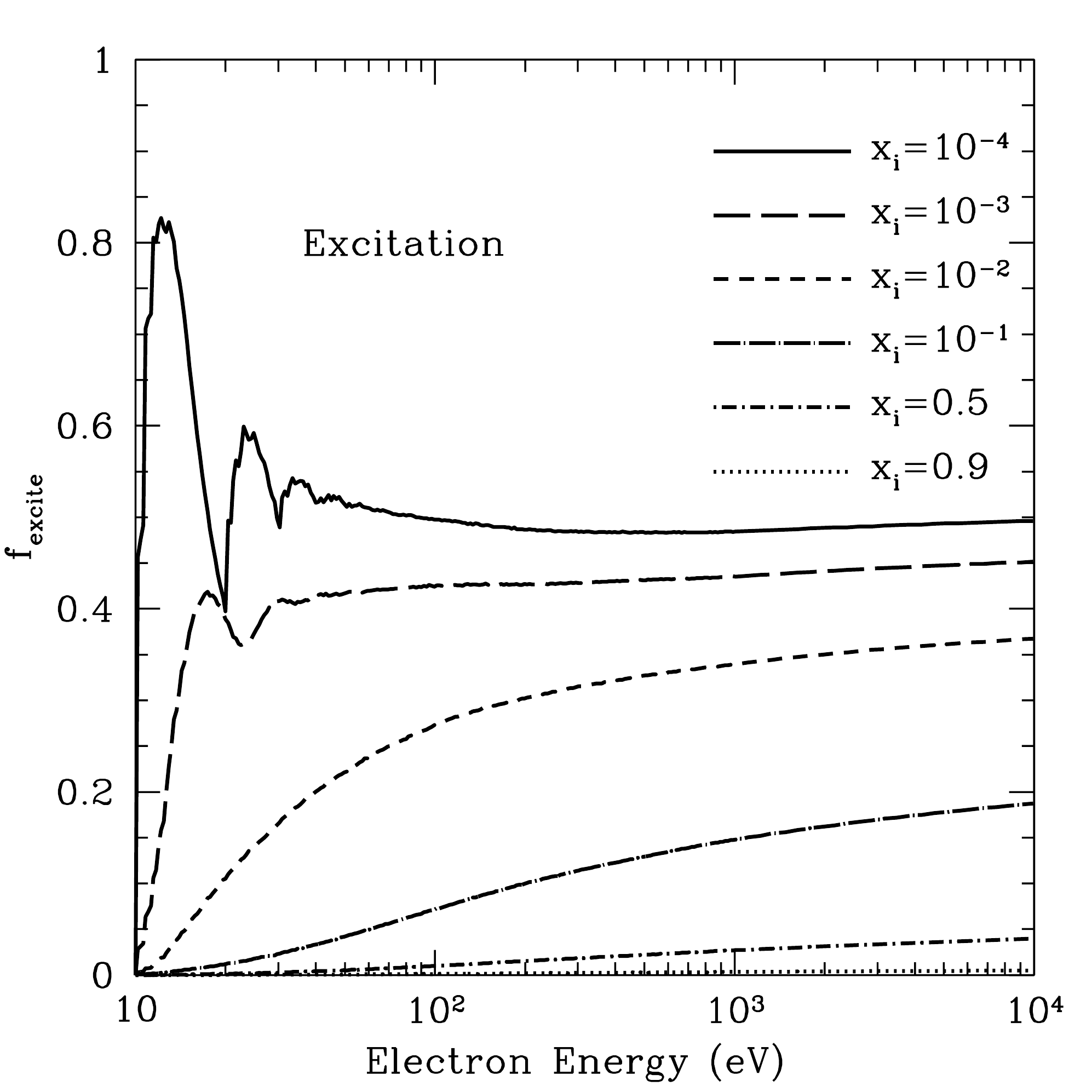} \includegraphics[width=0.4\textwidth]{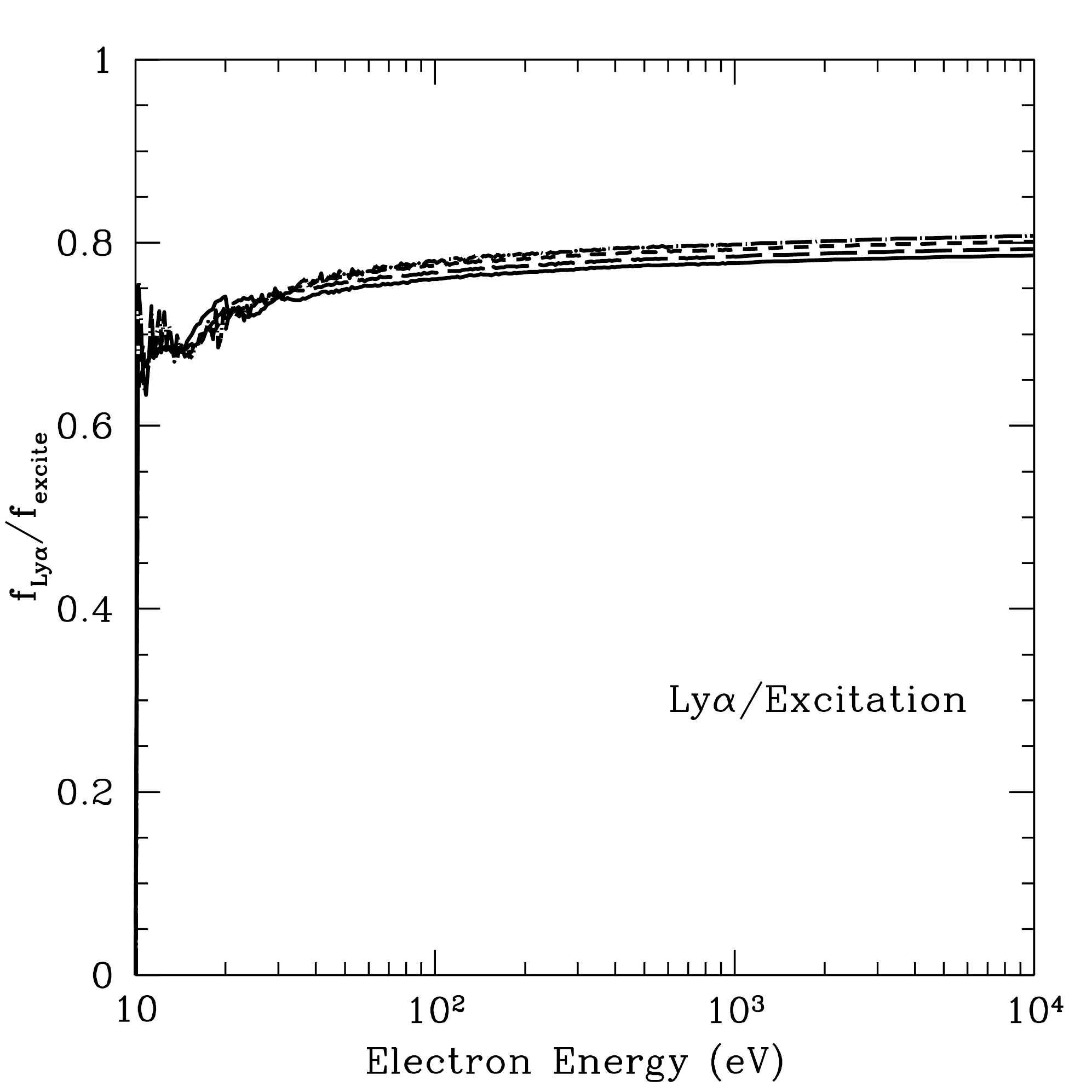}
\end{center}
\caption{Energy deposition from fast electrons. We show the fraction of the initial X-ray energy deposited in ionization (upper left), heating (upper right), and collisional excitation (lower left), as a function of electron energy and for several different ionized fractions $x_i$. The lower right shows the fraction of the collisional excitation energy deposited in the HI Lyman-$\alpha$ transition, $f_{\rm Ly\alpha}$.  Reproduced from S.~R. Furlanetto \& S. Johnson-Stoever, ``Secondary ionization and heating by fast electrons,Ó \emph{Monthly Notices of the Royal Astronomical Society}, vol. 404, pp. 1869-1878. Copyright OUP 2010.}
\label{fig:xrayheat}
\end{figure}

The details of this process have been examined numerically \cite{shull85,valdes08,furl10-xray}, and Figure~\ref{fig:xrayheat} shows some example results.\footnote{Note that these results are relative to the initial X-ray energy; some others in the literature instead use present results relative to the primary electron's energy.}  Note that the deposition fractions are smooth functions at high electron energies but, at low energies -- where the atomic energy levels become relevant -- can be quite complex. A number of approximate fits have been presented for the high-energy regime \cite{ricotti02, volonteri09}, but they are not accurate over the full energy range. A crude but useful approximation to the high-energy limit often suffices \cite{chen04-decay}:
\begin{eqnarray}
f_{\rm heat} & \sim & (1+2 x_i)/3 \nonumber \\
f_{\rm ion} \sim f_{\rm excite} & \sim & (1-x_i)/3,
\label{eq:fxapprox}
\end{eqnarray}
where $x_i$ is the ionized fraction. In highly ionized gas, collisions with free electrons dominate and $f_{\rm heat} \rightarrow 1$; in the opposite limit, the energy is split roughly equally between these three processes. However, the complexity of the behavior at low electron energies -- together with the increasing optical thickness of the IGM in that regime, and the fact that most sources are brighter in this soft X-ray regime -- suggest that a more careful treatment is needed for accurate work. \cite{furl10-xray} recommend interpolating the exact results.

\subsection{Other Potential Heating Mechanisms} \label{heat-other}

We close this section by noting that other heating mechanisms have been considered in the literature. One possibility is the heating that accompanies structure formation. When regions collapse gravitationally, they are heated by adiabatic compression (which we will discuss in Chapter 3), which is a minor effect. But, if the resulting gas flows converge at velocities above the (very small) sound speed, they can also trigger shocks, which convert a large fraction of that kinetic energy into heat. Analytic models and simulations suggest, however, that structure formation is still sufficiently gentle during the Cosmic Dawn that these shocks will have little effect on the 21-cm signal.  \cite{furl04-sh, kuhlen06-21cm, mcquinn12}. 

Finally, exotic mechanisms like dark matter annihilation or decay, primordial black hole emission, and other speculative processes can also affect the thermal evolution of the IGM during the Dark Ages. We will discuss such possibilities further in Chapter 3.

\bibliographystyle{plain}
\bibliography{ch2-refs}

\end{document}